\documentclass[prl,twocolumn]{revtex4}
\usepackage{epsfig}
\def\be{\begin{equation}}
\def\ee{\end{equation}}
\def\ba{\begin{eqnarray}}
\def\ea{\end{eqnarray}}
\def\bx{{\bf x}}

\begin{document}
\title{One-Way Hysteresis in the Depinning Transition}
\author{Ron Maimon$^1$ and J. M. Schwarz$^2$}
\address{$^1$ Gene Network Sciences, 2359 Triphammer Road, Ithaca, New York 
14850} 
\address{$^2$ Department of Physics, Syracuse University, Syracuse, 
New York 13244}
\date{January 23, 2003}

\begin{abstract}

We identify a mechanism for a type of hysteresis 
which we predict to occur in a variety of depinning transitions.
We show that the phenomenon of one-way hysteresis is generic to stress-overshoot models of the depinning transition,
and we show how to calculate the size of the hysteresis gap exactly in a large class of models.
Unlike thermodynamic phase transitions, the dynamic phase transition is still {\em continuous}
despite the presence of hysteresis because the terms which produce the hysteresis are 
renormalization group irrelevant. We discuss the experimental and numerical signatures of 
one-way hysteresis, which includes microscopic nucleation.

\end{abstract}
\maketitle

Many dynamical phase transitions display hysteresis in the transition from the static to
the moving phase. The most familiar of these is solid-on-solid friction-- the value of the
applied force which causes a solid to start sliding is greater than the force at which a
moving solid will stop. Another well studied example is the propagation of cracks in stressed solids.
The crack moves forward at a nonzero velocity when the force on the solid exceeds a critical value \cite{crack1},
and as the force is reduced, it is likely that the crack stops moving at a second critical force
lower than the first. However, the crack front posesses a well-defined roughness exponent in a variety of experiments \cite{rough}, which
suggests that the depinning transition is continuous.  

Hysteresis also occurs in other dynamical depinning transitions transitions: the current-flow transition in
charge-density wave solids \cite{cdw}, the depinning of vortex-lines in type II superconductors
\cite{vortex}, 
the onset of motion in hetereogeneous faults \cite{earth}, and 
the wetting  of a rough surface by helium-4 \cite{contact}. 
Previously it had always been assumed that the hysteresis implies that the transition is
discontinuous, by analogy with the world of thermodynamic phase transitions where hysteresis only
occurs in the first-order ones. We show that in nonequilibrium phase 
transitions this is not so, and that a particular one-way hysteresis of the type observed
naturally occurs in {\em continuous} phase transitions. This suggests that, among the
systems we observe to have one-way hysteresis, a number of them can still be described by a
statistical field theory at their critical point. We can show this because the
perturbation that generates one-way hysteresis is renormalization group
irrelevant, so the models with hysteresis remain in the same
universality class with the same critical exponents as in the absence of the 
perturbation.
The universal region is in a hard-to-reach corner of parameter space, however.  To get there,
we must first increase the applied force to start the system moving, and then decrease it until
the velocity is small. If a large fluctuation causes the system to stop, we must repeat the
procedure. This makes both numerical and experimental measurements of transition
quantities arduous.

The source of the hysteresis in our model is the nonadditive component of the transient stresses
in the system, and we can subtract it out by a simple procedure. The subtracted models are
equivalent to the unsubtracted ones, 
not only in the universal region but everywhere, except that the one-way hysteresis is removed.
The subtraction procedure is the main calculation tool, and it allows us to demonstrate that the
hysteresis gap exists, and to calculate the size of the gap in our models exactly.
Performing the subtraction makes numerical analysis of the transition much simpler, and
the subtracted models allow us to investigate the phase diagram of the continuum limit more
easily.

The models we analyze are minor generalizations of the stress-overshoot model introduced by Schwarz
and Fisher \cite{over1}, \cite{over2}.
Earlier, Fisher et al. \cite{fish} had suggested that models of this sort might possess a nonvanishing
hysteresis gap. 
Surprisingly, for the particular stress-overshoot model in \cite{over2}, this is not so---
the hysteresis vanishes in the infinite system limit. But the hysteresis is abnormally 
persistent, and this indicates that a small modification should produce true hysteretic behavior.

Our models have a finite hysteresis gap in the infinite system limit, proving
Fisher's conjecture. Furthermore, the one-way hysteresis occurs in the generic case---
earlier models had been accidentally fine-tuned so that the hysteresis was a finite-size
effect. But the existence of the gap is paradoxical-- the term that produces it is 
irrelevant in the renormalization group sense, and should not affect the critical behavior.
We resolve the contradiction by showing precisely how an irrelevant perturbation can
introduce one-way hysteresis without otherwise altering the critical behavior 
\cite{foot1}. The close
relation between the stress-overshoot models and our models provides an explanation for the
persistence of the finite-size hysteresis in \cite{over2}.

The model describes a surface whose position $h(\bx)$ increases by jumps. The
position of the surface describes the contact of two solids, or the position of a crack front, or the
phase in a charge-density wave. There are pinning forces which hold the surface back, and a global driving
force which pushes the surface forward. When the total force on a site is greater than the pinning force,
the site jumps. The sites communicate via an elastic force, which is linear, as if every site was
connected to its nearest neighbor by springs. In addition, there are transient velocity-dependent
forces on neighboring sites which vanish after a short period of time. These stresses are analogous
to inertial terms in highly dissipative equations, and like those terms they are generically irrelevant.
These are the stress overshoots which are responsible for the hysteresis.

The stress overshoots model a variety of very different physical effects. In a propagating
crack front, the stress overshoots model the effect of the acoustic wave which propagates in
the solid whenever a segment of the crack moves forward. This acoustic wave pushes on nearby
segments, possibly causing them to depin. In models of solid-on-solid friction, a stress
overshoot can model the transient component of the adhesive force between the solids, which rises
to a maximum value over a period of time.

To study these effects, 
on every site $\bx$ of a $d$-dimensional lattice define:\\
$h(\bx)$ -- the position (height) of a $d$-dimensional surface, initially zero.\\
$v(\bx)$ -- the velocity, either zero or one.\\
$f(\bx)$ -- a pinning force uniformly random in $[0,P]$.\\

There are three global parameters:\\
$P$-- the maximum strength of the pinning.  $P>2d$ to avoid lattice artifacts.\\ 
$F$-- the global applied force\\
$M$-- the magnitude of the stress overshoots.\\

To update the model:\\
1. Calculate the {\em elastic-stress} $E$ as the sum of $h({\bf y}) - h(\bx)$ over all nearest
neighbors ${\bf y}$.\\
2. Calculate the {\em overshoot-stress} $S$ as the sum of $v({\bf y})$ over all the nearest
neighbors ${\bf y}$ including $\bx$ itself in the sum \cite{foot2}. \\
3. At every site \bx   where $E(\bx) + M S(\bx) + F > f(\bx) $, set $v$ to one, elsewhere
set it to zero. \\
4. Add $v$ to $h$ at every site. Wherever $v$ is nonzero, the site is said to have {\em hopped}. \\
5. At every $\bx$ that hopped, generate a new pinning force $f$ in $[0,P]$.\\

This model is a discrete form of a standard model for the depinning transition. The
parameter we vary is the applied force $F$, and the order-parameter is the average velocity $\bar{v}$.
When no overshoots are present ($M=0$), it has a continuous transition at the critical force
$F_c$. This transition has been extensively studied, and the critical exponents have
been calculated to first- \cite{onutt1},\cite{natt1},\cite{onutt2} and second- \cite{wiese} order in epsilon.

The key to understanding the transition in the $M=0$ model is the no-passing rule of Middleton
\cite{nopass}. To state the no-passing rule, we must imagine that the random forces $f$ have been
pregenerated at the beginning of the simulation, and they are discovered as the surface moves forward.
The pregenerated force may be thought of as a random function of the height $h(\bx)$. The no-passing
rule states that when two surfaces $h_1(\bx)$ and $h_2(\bx)$ move in the same pregenerated pinning,
the surface which is initially behind will at no time overtake the surface that is initially ahead.

The proof is by induction in the time. Assuming that $h_1$ is ahead
of $h_2$ at time $t$, in order to determine whether $h_1$ is still ahead at time $t+1$ we only need
to examine the points where $h_1$ and $h_2$ coincide. At these points, the pinning force on $h_1$
and $h_2$ are equal because the pinning is a function of the height. The elastic force on $h_1$ is
greater than or equal to the elastic force on $h_2$ because the elastic force is an increasing
function of the position of the neighbors. Therefore, wherever $h_2$ jumps, $h_1$ jumps too, and
$h_2$ is not ahead at time $t+1$. 

Only assuming that the steady-state behavior of the surface is motion with a constant velocity, the
no-passing rule allows us to conclude that the velocity must be a single-valued function of the
applied force $F$, since at any force where two velocities may coexist, a surface moving with the
faster velocity will overtake a surface moving with the slower velocity, even if it is initially behind.
Since it is intuitively clear that the velocity is an increasing function of the applied force, the
no-passing rule also makes it all-but-certain that the transition in this model is continuous, since
at any jump-discontinuity the limits $\bar{v}(F+)$
and $\bar{v}(F-)$ are different, and we may start two surfaces at two infinitesimally different forces and
let the faster overtake the slower \cite{specialcase}. While these arguments as stated are not mathematically rigorous,
they are significant, because this transition is one of the few whose order may be established
theoretically.

From the proof, it is clear that no-passing is a generic feature of any elastic force law for
which the elastic force increases when the height of the neighbors increases. This implies that
adjusting the elasticity law in any reasonable way cannot produce hysteresis --- a new physical mechanism
is needed. Because the stress overshoots are transient, they can circumvent no-passing. A moving surface 
can move past a stationary one because the stationary one feels no overshoots.

Now we will demonstrate that the addition of stress overshoots of the type we have considered gives
rise to hysteresis. To make the proof transparent, we will introduce three simple models, one of which
has no hysteresis, and the other two which have a one-way hysteresis gap of exactly $M$ units of force. We
will show that they are all related by a subtraction procedure that removes $M$ units of one-way hysteresis.
Applying the subtraction procedure to our model we will arrive at a subtracted model which possesses a
hysteresis gap which is exactly $M$ units of force smaller.  Irrelevance of the perturbation
implies that there is no hysteresis at all in the subtracted model, from which we conclude that
our model has exactly $M$ units of hysteresis.

The first model we introduce has a stress overshoot which makes it trivial to calculate the
hysteresis gap. The model is defined with an overshoot stress $S(\bx)$
which is a global OR of all the Boolean velocities, that is, it is equal to $1$ if there is a single
nonzero velocity anywhere in the lattice, and it is equal to zero only if the 
surface is motionless. This is
a global perturbation of the model, and so we have no right to expect that the behavior is unchanged.
But the behavior of this model at any $M$ is obvious --- the overshoots simply shift
the critical force by $M$ units downward when the interface is moving, and they do nothing at all when
the interface is static. The effect of the global overshoot is trivial-- the model will
start moving at a critical force $F_c$ at a finite velocity, and it will stop moving at a critical
force $F_c-M$ where it will display identical behavior (site by site and hop by hop) to the $M=0$
model. The transition remains continuous and the critical exponents are unchanged.

The second model we introduce is local, and is superficially different from the first. In this model,
which we call the {\em nonadditive model}, the overshoot stress is a local OR of the
neighboring sites' velocities-- $S(\bx)=1$ only if $v(\bx)$ is nonzero or $v({\bf y})$ is nonzero for
one of the nearest neighbors ${\bf y}$ of ${\bf x}$. But difference is superficial-- the nonadditive
model is (site by site and hop by hop) equivalent in the steady state to the global
model. In particular, it has a continuous phase transition and a one-way hysteresis gap of width
exactly $M$.

To prove this assertion, start the global model moving, and let it reach a steady-state at a nonzero
velocity. Whenever a site hops in the steady state, its local environment has necessarily changed on
the previous time step-- either it has itself hopped on the previous step, or one of it's nearest neighbors
have. Conversely, any site whose nearest neighbor hasn't hopped cannot hop, despite 
the global stress overshoot. So we may apply the overshoot stress only to the sites whose environment
has changed-- to those sites that hopped on the previous step and their nearest neighbors-- with no change
in the dynamics, and this is the definition of the nonadditive model. This completes the proof.

The nonadditive overshoot model is a local irrelevant perturbation of the $M=0$ model, which
introduces precisely $M$ units of one-way hysteresis and does nothing else. It
does not change the order of the transition or the critical exponents.

The subtraction procedure is inducted from the two models defined above. To find a model with no one-way
hysteresis from a given model, remove the non-additive component of the stress overshoots. The nonadditive
component is the minimum value of stress overshoot on any of the sites that are capable of hopping.
Since motion propagates locally from site to site, any model in which the stress overshoots are positive
and propagate faster than the elasticity will have one-way hysteresis of the type described. 
For our specific model at nonzero $M$, we produce the {\em subtracted model} by subtracting the
nonadditive component of the stress overshoot. In the subtracted model, the overshoot stress $S(\bx)$
is zero if zero or one neighbor hops, and $S(\bx)=n-1$ if $n$ neighbors of a site hop (recall that a
site is its own neighbor). To recover our original model, we must add $S_2(\bx)$-- a nonadditive stress
overshoot. In other words, $S_2(\bx)$ is the Boolean OR of all
$2d+1$ neighboring velocities. Because in steady-state every site that hops has a neighbor that hops,
the only effect of $S_2$ is to add one-way hysteresis of gap size exactly $M$ \cite{meanfield}.

The subtracted model is a local perturbation of the $M=0$ model, and the perturbation is necessarily
irrelevant in a renormalization group sense, since it is local and breaks no symmetry.
We conclude that any hysteresis in the subtracted model is a finite-size effect in this regime, a result
confirmed by numerical simulation (see Fig. 1).

\begin{figure}[t]
\begin{center}
\epsfxsize=8.0cm
\epsfysize=8.0cm
\epsfbox{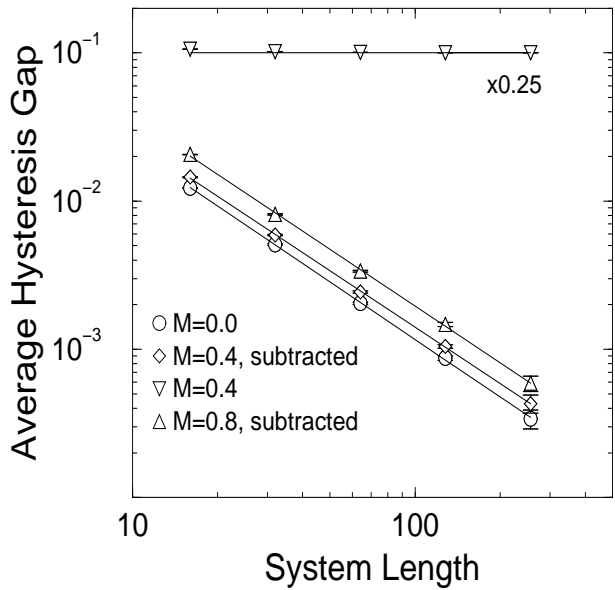}
\caption{Plot of the finite-size hysteresis gap for the stress overshoot and 
subtracted models.  
  To generate the data, we start the system in the moving phase and 
lower $F$ until the surface stops. We then increase the $F$ 
back up again until the surface begins to move. The symbols denote the data, 
and the curves denote the least-squares fits. 
The error-bars indicate $1\sigma$
statistical uncertainty.  When $M=0.0$, we expect the gap to 
vanish as a power law in the system length; the exponent is 
$1/\nu=4/3$ to first-order in perturbation theory [13,14], 
where $\nu$ is the correlation 
length exponent.   
We measure $1/\nu=1.29\pm
0.03$.  For the subtracted model, when $M=0.4$, $1/\nu=1.26\pm 0.03$ and 
for $M=0.8$, 
$1/\nu=1.27\pm 0.03$. With the subtracted model, there is an $M$-dependent coefficient that increases 
with increasing $M$. For the stress overshoot 
model with $M=0.4$, the gap is 
$0.4$ within one $\sigma$ at least for the 
larger sizes. We have multiplied this data by $\frac{1}{4}$ for plotting 
purposes.  For all data, $d=2$ and $P=4$ [19].}
\end{center}
\end{figure}

Any model which propagates motion from one site to another and which includes a stess-overshoot
will generically display one-way hysteresis. The appropriate subtracted model removes the minimum
stress overshoot which is applied to any moving site.
The remaining perturbation is irrelevant, and does not change the critical behavior for small values
of $M$. Since the phenomenon is so general, we believe that it explains the origin of one-way hysteresis in
at least some of the systems that display it.
 This implies that one-way hysteretic transitions are still continuous, despite 
appearances.

Simulating our original model and the subtracted model at large values of the parameter $M$,
we find that there is a tricritical point at a finite value
$M_c$ of $M$, beyond which the hysteresis changes character. For $M>M_c$, subtraction no longer
removes the hysteresis, and the transition is truly discontinuous. At $M=M_c$, the universality
class of the transition is changed. The tricritical point is a generic feature of overshoot
depinning models.

There are two peculiar properties of the hysteresis in the unsubtracted models.
First, the nucleation radius is not infinite as it is in first-order thermal transitions. If one
well-placed site were to hop when the applied force is in the hysteresis gap,
the whole surface would start to move.
Second, the hysteresis loop does not vanish for large systems as the inverse logarithm
of the system size. When the surface is static, the probability of nucleation is
zero because there is no dynamics. If we were to add a small amount of spontaneous hopping,
the hysteresis loop would vanish as a power of the system size and the time.
In first-order transitions, the nucleation bubbles becomes larger and exponentially more
improbable as the critical point is approached; here, a nucleation bubble may be microscopic.
It stays microscopic even at the critical transition.

The characteristics of one-way hysteresis are:\\
1. The stationary phase is unstable.\\
2. The nucleation radius is finite at the transition point\\
3. With local fluctuations, the probability of nucleating a bubble vanishes as a {\em power}
of the system size like a finite-size effect, not logarithmically like a first-order hysteresis
loop.\\
4. The stable phase is described by a continuum theory at the critical point.\\
\\
The one-way hysteresis which we have seen occurs is, in some sense, a phony hysteresis. The reason is that
it is only present because when the interface is static, all overshoots are absent and do not
contribute to the motion. If we were to add a small amount of spontaneous hopping, the
hysteresis gap would vanish as a power of the system size. The only reason we 
may observe this effect 
is because in many depinning transitions thermal fluctuations may be ignored.

An earlier model \cite{over2} included stress-overshoots only on nearest neighbors, but did not include
overshoots on the site itself. Because of this, the nonadditive component of the stress overshoots
is zero. If a small amount of self-stress overshoot is added, the system acquires a small amount
of one-way hysteresis. Without the self-stress, a site which hops twice in a row without being triggered by its nearest-neighbors on the second hop does not feel the stress
overshoot the second time around. The site then has a chance to land and be pinned very weakly,
and a weakly pinned site may serve as nucleation centers when the force is increased
in the static phase. The result is that the hysteresis gap vanishes as the
system size grows, albeit slowly because the nucleation centers are so rare. The interesting
observation is that the nucleation centers are one (well-placed) site wide.
This is the numerical and experimental signature of phony hysteresis. A scattering of
microscopic seeds will destroy it. In solid-on-solid friction, for example, a sound wave with
a fixed frequency and amplitude should destroy any phony hysteresis,
while it would only reduce, but not eliminate, the hysteresis in a true first-order transition
such as the one that occurs for $M>M_c$. 

For completeness, we note that when $S(\bx)=v(\bx)$, that is, when a site only applies a stress
overshoot to itself, there is no hysteresis of any kind for any value of $M$ no matter how large.
This is an accident of the parametrization of the lattice model. If the overshoot stress is decays
exponentially over many time steps, for sufficiently slow decay rates we approach the tricritical
point in these models as well. There is no one-way hysteresis
in this model of course, because the nonadditive stress overshoot is zero, since the overshoot
stress is not applied to the neighbors.

The authors would like to thank Daniel S. Fisher and Alan Middleton 
for very useful discussions, 
and Alan Middleton for comments on an earlier manuscript.
R.M. is supported by Gene Network Sciences Inc.
J.M.S. is supported by NSF via grants DMR-0109164, DMR-9805818.


\begin{thebibliography}{99}
\bibitem[10] {foot1} These are not ``dangerously irrelevant'' variables.
\bibitem[11] {foot2} The model defined in \cite{over2} only differs in
that it does not apply the overshoot stress to the site itself. 
Since the overshoot only acts for one
time step and very few sites hop twice, the difference between our model and the previous one is only noticeable
after very rare events.
\bibitem[19] {foot}  For the simulations, we initialize $h$ uniformly 
random on the interval $[0,1]$.  Then $P$ can take on any value without 
causing lattice effects.
\end{thebibliography}
\end{document}